\documentclass[12pt]{iopart}

\usepackage{iopams}
\usepackage{graphicx}  
\begin{document}

\title{Casimir forces from a loop integral formulation}
\author{James Babington}

\address{
Quantum Optics and Laser Science, Blackett Laboratory, \\
Imperial College London, Prince Consort Road, \\
London SW7 2AZ, U.K.}
\ead{j.babington@imperial.ac.uk}

\begin{abstract}
We reformulate the Casimir force between bodies in non-trivial background media. The force may be written in terms of loop variables, the loop being a curve around the scattering sites. A natural path ordering of exponentials takes place when a particular representation of the scattering centres is given. The basic object to be evaluated is a reduced (or abbreviated) classical pseudo-action that can be operator valued, and can be obtained from a classical path integral description.

\end{abstract}


\section{\label{sec:INTRODUCTION}Introduction}

In a recent paper~\cite{babington-2009}, Casimir forces were calculated for a collection of dielectric spheres in a constant background dielectric media. By evaluating the stress tensor compatible with the Lorentz force law~\cite{raabe:013814,scheel-2008-58}, it was possible to calculate the force for the case of two and three sphere systems as a multiple-scattering expansion. This approach is indeed very similar to the scattering methods used recently to evaluate functional determinants which give the interaction energies amongst bodies~\cite{rahi-2009,Emig:2007cf,Bulgac:2005ku,Kenneth:2006vr,Lambrecht2006}. Certain aspects of the scattering approach and the use of functional determinants were also present in the earlier work~\cite{Feinberg:1970zz, renne71,Balian:1977qr,Bordag:1983zk}. The ground state energy of the system is projected out from an effective action which can be written as a functional integral over the physical fields. For an overview of these methods see the recent book~\cite{advancescasimir}. Both approaches share in common the formulation of translation matrices that map eigenfunctions between scattering centres and the corresponding scattering coefficients for each sphere. The approach in~\cite{babington-2009} differs from the functional determinant approach principally because it uses macroscopic quantum electrodynamics~\cite{scheel-2008-58} to formulate the problem. In this framework, quantisation in real media is explicitly performed. Using this method one evaluates a force directly from a stress tensor; and it uses the fluctuation-dissipation theorem to evaluate the two-point function of the noise currents (which in turn are used to evaluate the bilinear correlation functions of the physical electromagnetic fields~\cite{scheel-2008-58}). All of the quantum mechanical content of the problem resides in this two-point current correlator giving it as an order $\hbar$ expression. One calculates the physical observable (i.e. the force) which must be free of divergencies from the outset. Whilst the two approaches will agree for bodies in vacua to leading order, this will not be the case for bodies in real magneto-dielectric media. 

In this paper we study some properties of the force on a general submanifold by reformulating the expansion in terms of a set of loop variables. The observation in~\cite{babington-2009}, that the total path length of the particular set of scattering events plays a key role in evaluating the frequency integral, leads naturally to the concept of the loop integral of all the scattering events. This is because in the last stage of the calculation when evaluating the force on a body or at a point, we must set the arguments of the Green's function to coincide, thus closing the path to form a loop. It can be thought of heuristically as a causally propagating photon scattering off the spheres and finally arriving back where it started. In a similar way, this is basically what is done when one attempts to evaluate the density of states. Loop variable formulations are important in other areas of physics, notably lattice gauge theory where the Wilson loop~\cite{makeenko-2009} plays a ubiquitous role, and loop quantum gravity~\cite{Gambini:1996ik}, where the definition of a loop derivative also appears.

It is possible to further rewrite this simply in terms of a loop integral of the conjugate generator of translations and the potential that captures the system (e.g. step functions representing spheres, smooth short distance or singular type potentials). Note that this is not the momentum, but rather the generator of translations (with dimension inverse length), a purely classical operator resembling a covariant derivative and therefore \emph{without} Planck's constant or a mass scale entering at all. Loop integrals have appeared before in~\cite{PhysRevA.58.935,schaden:042102} where Casimir energies on different spaces are approximated from a semi-classical description (for a description of the density of states in terms of Maslov indices see~\cite{PhysRevA.42.1907}). This is, practically speaking, equivalent to the optical approximation~\cite{Scardicchio2005552,scardicchio-2006-743,schroeder-2005-72} in evaluating the Casimir energy and the inclusion of closed orbits as opposed to only the periodic orbits in~\cite{PhysRevA.58.935,schaden:042102}. These works share in common a saddle point approximation of some functional integral. Complimentary ideas are found in the worldline formalism~\cite{Gies:2001tj} applied to scalar fields that enables Casimir energies to be calculated numerically. For applications of these methods see~\cite{scardicchio-2006-743,schroeder-2005-72,Gies:2001tj,gies-2003-018,gies-2006-96, gies-2006-74}.

Looking further ahead, we might expect this potential is related to some background 3-metric or affine connection that depends on frequency, since the response functions are non-local in time but may be local in space. Thus an expression similar to the adiabatic invariants and WKB/Sommerfeld quantisation may be anticipated, except that this is purely classical and contains the length of the loop in an implicit manner. For a related earlier discussion of such integrals that are not periodic and turn stochastic see~\cite{Zaslavskii77}.

In principle the idea presented here is an exact method; one uses basically the Fradkin representation~\cite{Fradkin:1966zz, FriedBook,Kleinert:2004ev} of the Helmholtz operator for the physical fields which can then be turned into a particle type path integral. To then evaluate this requires either very simple media or approximate techniques. They are expected to coincide with those found in~\cite{PhysRevA.58.935,schaden:042102, Scardicchio2005552} when they are approximated, for example by straight line rays, and one could imagine that they provide a useful 'discrete' way of realising them.

The outline of this paper is as follows. In Section~\ref{sec:Generalities} we give a description of the field equations we are considering and their stress tensors. In Section~\ref{sec:Loop} we show how the $N$-sphere Casimir force can be reformulated as a loop integral, together with some details of the classical generator of translations and the use of representations. We also discuss the overlap with a path integral representation. In Section~\ref{sec:Examples} we give two examples; electromagnetic fields in a magneto-dielectric background in three dimensions and a scalar field in two dimensions. Finally in Section~\ref{sec:Conclusions} we conclude.

\section{Some generalities}
\label{sec:Generalities}
\subsection{Equations of motion}

We consider some general background media described by a set of potentials, through which a particular spin state propagates. The three most usual are spin $0,1/2,1$ (scalar, fermion, vector), with equations of motion given by
\begin{eqnarray}
(\nabla^2 +k^2)\phi= j_N^{(0)}, \\
i\gamma^i(\partial_i +k_i )\psi=\mathcal{J}^{(1/2)}_N, \\
\nabla \wedge \mu^{-1} \cdot \nabla \wedge \mathbf{E} -(\omega^2/c^2)\epsilon \cdot \mathbf{E}=i\omega \mathbf{J}^{(1)}_N, \\ 
\nabla \wedge \mathbf{E} +i\omega \mathbf{B}=0.
\end{eqnarray}
To evaluate the force on some object (sub-manifold $\mathcal{O}$) we need to construct a stress tensor out of the physical fields and integrate this over the the volume of the submanifold
\begin{equation}
F^i(\mathcal{O})=\int_{\mathcal{O}}d^nx  (\nabla_jT^{ji}).
\end{equation}
If the object is compact then we can calculate a force as opposed to a force density when one or more dimensions of the subspace are non-compact. For the three most relevant cases, the stress tensors one can consider are
\begin{eqnarray}
T^{ scalar}_{ij}&=&\partial_i\phi\partial_j \phi , \\
T^{ spinor}_{ij}&=&\frac{i}{2}\bar{\psi} \gamma_{i}\stackrel{\leftrightarrow}{\partial}_{j} \psi , \\
T^{ vector}_{ij}&=&\mathbf{E}_i\mathbf{E}_j+\mathbf{B}_i\mathbf{B}_j -\frac{1}{2}\delta_{ij}(\mathbf{E}\cdot\mathbf{E}+\mathbf{B}\cdot \mathbf{B}).
\end{eqnarray}
At this point the classical field equations need to be solved subject to the usual continuity and boundary conditions. The bilinear physical fields in the above can be written in terms of classical Green's functions together with the appropriate noise currents that persist in the background. Using the fluctuation-dissipation theorem, the vacuum expectation values of the two-point current correlation functions can be evaluated thus giving an explicit expression for the stress tensor.

\section{The loop integral variables}
\label{sec:Loop}
\subsection{A loop integral for the inter-sphere force}

We start with the expression given in~\cite{babington-2009} for the Casimir force on a sphere due to multiple-scattering for the collection of $N$-spheres (all are dielectric). See~\cite{rahi-2009} for a related discussion in terms of energy functionals. The physical dispersion force carriers are the electric and magnetic fields. For all the diagrams of order $N$ in the number of interactions, the $N$-body force was found to be (where the frequency is rotated to the imaginary axis $\Omega=i\omega$)
\begin{eqnarray}
\label{eq:NSPHEREFORCE1}
\mathbf{F}[1|N-1]=-(-1)^N\frac{\hbar }{4\pi } \int_{0}^{\infty}d\Omega  \cot \left(\frac{\hbar \Omega}{k_BT}\right)\cdot \langle\mathbf{1}|  [ \alpha^{1} (\Omega R[1]) \nonumber \\
 \sum_{i,j,\cdots, m=2 }^{N}A^{1,i}(\mathbf{r}[1,i])\alpha^{i} (\Omega R[i]) 
A^{i,j}(\mathbf{r}[i,j])\alpha^{j} (\Omega R[j])\cdots \nabla_{\mathbf{r}[m,1]}A^{m,1}(\mathbf{r}[m,1])] \nonumber \\
\cdot (k R[1])\cdot j(R[1])\cdot n(R[1])W(\Omega R[1])|\mathbf{1}\rangle \nonumber \\
\equiv  -(-1)^N\frac{\hbar }{4\pi }\sum^{N}_{i=2} \nabla^{last}_{\mathbf{r}[i,1]}\int_{0}^{\infty}d\Omega  \cot \left(\frac{\hbar \Omega}{k_BT}\right) \mathcal{Z}[\alpha,A,W].
\end{eqnarray}
Briefly the terms occurring in Equation~(\ref{eq:NSPHEREFORCE1}) are : the difference vectors $\mathbf{r}[i,j]$ connecting each of the sphere centres; the Mie scattering coefficients of each dielectric sphere $\alpha^{i} (\Omega R[i])$; the translation matrices $A^{i,j}(\mathbf{r}[i,j])$ that map TE and TM modes from one sphere origin to another; the $SO(3)$ vectors $|\mathbf{1}\rangle$ that truncate the matrix representations to some finite order in the angular momentum $L$; the derivative operator $\nabla^{last}_{\mathbf{r}[i,1]}$ that acts on the last translation matrix of the order $N$ polynomial; and the function $W$ is a measure due to the two separate contributions in the stress tensor, arising from the individual field components $\mathbf{E}_m\mathbf{E}_n+\mathbf{B}_m\mathbf{B}_n$ and the part where traces have been performed $\delta_{mn}(|\mathbf{E}|^2+|\mathbf{B}|^2)$. Of course, this could have equally well been done for spin-0 or spin-1/2 massless fields that are then the effective force carriers. Obviously the total force requires adding up all the lower order disconnected diagrams as well. 

It can be seen from Equation~(\ref{eq:NSPHEREFORCE1}) that the central object of any Casimir interaction calculation is the evaluation of the quantity $\mathcal{Z}[\alpha,A,W]$. In fact as will be shown below, this is no more than a specific \emph{representation} of an object that shares some structure with semi-classical physics. It is clearly some type of loop integral, as we start and finish on the same point of the particular sphere where the force is being evaluated. We can write this in a shorthand notation by the introduction of a path ordered exponential (see~\cite{makeenko-2009} for details of how this is used in gauge theory) and replacing the truncation vectors $|\mathbf{1}\rangle$ with a trace over $SO(3)$ indices (up to some integer $L_{max}$), together with retaining only the polynomials in the translation matrices that form the closed loop index wise
\begin{equation}
\label{eq:pathorder}
\mathcal{Z}[\alpha,A,W]= \mathbf{ Tr}\left[ \mathcal{P} \exp \left( \sum_{i=1}^{N}\sum_{j\neq i}^{N}\alpha^{i} (\Omega R[i])A^{j,i}(\underline{r}[j,i]) \right)\right]^{loop(A)}_{L_{max}}.
\end{equation}
This means that the first contribution retained is the second order contribution in different scattering coefficients (the familiar two body forces). The necessity of path ordering in an expression like this is due to the fact that in general we are dealing with non-commuting matrix representations. Thus to form a continuous path  we must specify the order of the translation matrices so that they are sewn together in the correct way. Indeed, Equation~(\ref{eq:pathorder}) is really defined by the power series expansion, as the exponent is not meaningful for a set of translation matrices of a different dimensions (for example a dipole quadrupole interaction). The suggestive feature of this form however, is that we encounter a closed loop of finite distance translations in an exponential similar to what is encountered in semi-classical physics.

\subsection{The translation operator}

We now try to develop the loop integral formulation and see the connection with path integral methods. Define the translation operator
\begin{equation}
U(y,x;\hat{D}):= \exp\left[ {\int^y_x dz^i \hat{D}_i(z)}\right],
\end{equation}
where $\hat{D}_i$ is the classical generator of translations in a given background. The translation matrices that map the vector wave-functions between different scattering centres~\cite{Mackowski06081991} are then just representations of this operator in the $SO(3)$ TE/TM basis. For example, the $A$ matrices used in~\cite{babington-2009, Mackowski06081991} are just the constant finite displacement versions of this operator (the background is constant, the spheres are viewed as perturbing potentials, and $\hat{D}_i=\hat{\nabla}_i$ is the standard derivative operator)
\vspace{0.5cm}
\begin{equation}
\label{eq:Arep}
\fl A^{L2,m2}_{L1,m1}(k|y-x|)=\frac{\int_{S^2}\langle L2,m2,h^{-}_{L2}(k|x|)|\mathbf{L}_i \cdot U(y,x;\hat{D})\cdot \mathbf{L}_i |L1,m1,h^{+}_{L1}(k|x|) \rangle}{[\int_{S^2}|\langle L2,m2,h^{-}_{L2}(k|x|)|\mathbf{L}_i|^2]^{1/2}\cdot  [\int_{S^2}|\mathbf{L}_i |L1,m1,h^{+}_{L1}(k|x|) \rangle|^2]^{1/2}}.
\end{equation}
\vspace{0.5cm}
The expression we have encountered for the force in Equation~(\ref{eq:pathorder}) can be partially recast using Equation~(\ref{eq:Arep}). By using the orthogonality of the vector eigenfunctions, we see that all we are doing is furnishing a representation of the translation operator by the insertion of complete sets of states when we have perturbing scatterers  present. 

More generally, if we were now to change from the step function potentials relevant to the hard sphere scattering (together with the application of continuity equations across the surfaces) to a general 'softer' set of potentials and consider $\mathcal{Z}$ at a particular point (or on a submanifold), then we would not be dealing with straight line rays and reflection coefficients, but rather curved rays in some effective potential. This is a further prompting to write the fundamental objects in terms of loop variables. For a spin-$s$ field we write the corresponding loop integral as
\begin{eqnarray}
\label{eq:LOOP2}
\mathcal{Z}^{(s)}[\Omega,x] &=& \langle\mathbf{F}^s_x|  \exp \left[-\oint_{\mathcal{C}_x}dq^i\hat{D}_i(q,\Omega) \right] |\mathbf{I}^s_x\rangle \\
&\equiv&\mathbf{ Tr}\left(  \exp \left[-\hat{S}_R(\mathcal{C}_{x}, \Omega) \right] \right)_{\{\mathbf{I},\mathbf{F},x,s\}},
\label{eq:LOOP3}
\end{eqnarray}
where $x$ is the initial and final point of the loop, and $|\mathbf{I}\rangle$ and $|\mathbf{F} \rangle$ are the initial and final states (which may differ for example by the boundary conditions imposed on the eigenfunctions used to represent in and out modes). The generator $\hat{D}_i$ of translations and the loop $\mathcal{C}_{x}$ implicitly depend on the background potentials. In the next subsection we shall show how this loop integral arises in a path integral description. The canonical generator of translations $\hat{\nabla}_i$  (i.e. the gradient operator for smooth functions) is promoted to a constrained covariant derivative operator in general
\begin{equation}
\label{eq:generator}
\hat{\nabla}_i \rightarrow \hat{D}_i.
\end{equation}
In Equation~(\ref{eq:LOOP3}) we encounter for the first time the \emph{reduced pseudo-action} $\hat{S}_R$ which is a dimensionless object, since it is defined using the generator of translations rather than the canonical momentum. It is in general operator valued, being realised on eigenfunctions of Helmholtz type operators. Obviously it is a functional of both the background effective potential and the loop. This should be contrasted with the semi-classical evaluation used in~\cite{PhysRevA.58.935,schaden:042102} where a saddle point approximation is used to evaluate the density of states. It is also different from the optical approximation~\cite{Scardicchio2005552}, where straight line elements dominate the path integral (giving a stationary phase) of a classical particle action with the mass scale set equal to unity. Since the loop integral developed above uses the fluctuation dissipation theorem to evaluate the quantum part, the physical fields 'momentum' density will diminish with subsequent windings of the loop, thus defining part of the perturbation theory. The tangent ray in contrast does not have this dynamical information encoded. Rather than considering periodic orbits in the usual particle phase space, one would expect the configuration space of windings is the appropriate one for dissipative backgrounds. See also~\cite{Zaslavskii77} for earlier work on non-periodic orbits in which trajectories become stochastic.

The last part to fully specify the $\mathcal{Z}$ function is in finding a set of loops in the background that describes how the force carriers propagate. This, together with the corresponding representations of the translation operator (if it necessary to cut the loop integral up) are the remaining quantities required in any evaluation. The assumption we now make in order to calculate the loop equation is that the background effective potential can be turned into an affine \emph{connection} with which to define parallel transport.  Thus the loop is given by the parallel transport of the loops tangent vector based on a connection derived from the background
\begin{equation}
\label{eq:geodesic}
\frac{d}{d\tau} \dot{q}^i +\Gamma^{i}_{jk} \dot{q}^j\dot{q}^k  =0.
\end{equation}
The significance of this statement is that it is only when the background potential possesses some type of critical point, singularity or sharp boundary that the loop integral can be non trivial. Equation~(\ref{eq:geodesic}) needs to be solved subject to the boundary conditions $q^i(initial) =q^i(final)$ and $\dot{q}^i(initial) =\dot{q}^i(final)$. It thus tells us the shape of the loop rather than its length, since we are not initially assuming a metric connection. In this manner once a connection has been derived, one is able to work out the space of loops (the set of closed paths labelled by $\tau\in [0,1]$)
\begin{equation}
\label{eq:Loops}
\mathcal{C}_x = \{ q^i(\tau)|\; \ddot{q}^i +\Gamma^{i}_{jk} \dot{q}^j\dot{q}^k  =0, 
 \; q(0)=q(1)=x,\; \dot{q}(0) =\dot{q}(1)\},
\end{equation} 
which in turn can be used to then evaluate the loop integral. Note that this prescription is different from choosing the initial and final momenta to coincide due to the dissipation in the background. Here instead we make only a statement about the shape of the loop rather than the dynamics.
 
The total loop integral must take into account all possible interactions which is reformulated into all possible periodic windings of the loops around the critical points (or non trivial features) of the background. One might think that there are a large space of such loops with equivalence under smooth deformations (i.e. not crossing critical points boundaries, etc.). The use of a connection to give a geodesic equation has the effect of collapsing this space down to just the one element for a given frequency. The final expression we find for the total $\mathcal{Z}$ function is
\begin{eqnarray}
\label{eq:Loop3}
\mathcal{Z}^{(s)}[\Omega,x]& = &\sum_{\mathcal{C}_x} \langle\mathbf{F}^s_x|  \exp \left[-\oint_{\mathcal{C}_x}dq^i\hat{D}_i(q,\Omega) \right] |\mathbf{I}^s_x\rangle \nonumber \\
&=& \sum_{\mathcal{C}_x}\mathbf{ Tr}\left(  \exp \left[-\hat{S}_R(\mathcal{C}_{x}, \Gamma, \Omega) \right] \right)_{\{\mathbf{I}^s_x,\mathbf{F}^s_x\}}
\end{eqnarray}
and the sum over loops can be expanded (for example in terms of winding numbers). This expression captures the essential elements of the classical scattering of the physical fields, together with Equation~(\ref{eq:NSPHEREFORCE1}) where the derivative acting on it removes the non interacting pieces.

\subsection{The relation to the Green function}

The loop integral we have derived may be thought simply to be the Green's function evaluated at coincident points. This is not quite the case as can be seen by considering a path integral representation (see~\cite{Kleinert:2004ev}) specifically for the electromagnetic fields. We consider now a more general background media given by the dielectric permittivity $\epsilon_{ij}(x,\Omega)$ and magnetic permeability $\mu_{ij}(x,\Omega)$ tensors, i.e. they have no spatial dispersion and are just local functions of position. Rewriting the field equation for the electric field as
\begin{equation}
\label{eq:Efe}
\left(\frac{c^2}{\Omega^2}\epsilon^{-1} \cdot \nabla \wedge \mu^{-1} \cdot \nabla \wedge +\mathbf{1}\cdot \right) \mathbf{E}(x,\Omega)=\left(\frac{c^2}{\Omega} \right) \epsilon^{-1} \cdot \mathbf{J}^{(1)}_N(x,\Omega),
\end{equation} 
we see the natural appearance of a dimensionless differential operator on the left. This operator has the same eigenfunctions and eigenvalues as the standard Helmholtz operator for the electric field. It can be used in the construction of propagators, whilst a modified evaluation of the noise currents two point function can be found to evaluate the electric field two point function. The corresponding two point function can be represented as (promoting the partial derivatives to covariant derivatives with connections based on the permittivity and permeability)

\begin{eqnarray}
\label{eq:twopointf}
\mathbf{\Delta}(z,y)&=&\langle z |\left[-c^2/\Omega^2 \nabla_{\epsilon}\wedge\nabla_\mu \wedge -\mathbf{1} \cdot \right]^{-1}| y \rangle \nonumber \\
&=& \int_0^{\infty}d\tau \langle z |e^{-\tau [ c^2/\Omega^2\nabla_{\epsilon}\wedge\nabla_\mu\wedge +\mathbf{1}\cdot]}| y \rangle.
\end{eqnarray}
By introducing an integration measure over paths connecting the two point this can be written as (where for simplicity we take the scalar version of this Helmholtz operator containing the two potentials)
\vspace{0.5cm}
\begin{equation}
\label{eq:twopointpi}
\fl \Delta(z,y) = \int_0^{\infty}d\tau \int^{z}_{y} [dq(t)]\sqrt{\det [(\Omega^2/c^2)(\epsilon\cdot\mu)]}\exp\left[-\int^{\tau}_0 dt (\Omega^2/c^2)(\epsilon \cdot \mu)_{ij} \dot{q}^i\dot{q}^j +1\right].
\end{equation}
\vspace{0.5cm}
Considering now the loop integral version $z=y$ for the closed orbits it is obviously necessary to compactify the real line $\tau\in S^1$ to the reduce the integral to a sum over windings, together with a specification of the integration measure. This is similar to the path integrals used in~\cite{PhysRevA.58.935} and~\cite{Gies:2001tj} where a fictitious time is introduced as well as mass scale set equal to unity and an appearance of Planck's constant in the particle action. As pointed out in~\cite{Scardicchio2005552}, it is misleading to call this a semi-classical evaluation because of the absence of the dimensionful Planck's constant. Concurrent with this is the absence of a mass or length scale with which to define a dimensionless action. The choice we have made in the above can be seen to be dimensionally correct and relevant at the local level. It useful to compare this expression with the first order form of the quantum path integral for a relativistic particle~\cite{Kleinert:2004ev} in a background metric:-
\begin{eqnarray}
 \Delta(x,x) &:=  \langle x|\frac{1}{\hbar^2\partial_{\mu}\partial^{\mu}+m^2c^2} |x\rangle \nonumber \\
 &= \int^{\infty}_{0} d\tau \oint_{q[0]=q[1]=x} [dq]\sqrt{\det(g)}\int [dp/\hbar] [de] \nonumber \\
 &\cdot \exp \left(-i/\hbar \int^{\tau}_{0}dt[ \dot{q}p+e(p^2+m^2c^2)]\right). 
\end{eqnarray}
Performing the integral over momentum, we get back to the configuration space (a proper treatment also requires gauge fixing the world line metric $e=1$ and the introduction of ghosts). It has the feature of inverting the background potentials used to define the Hamiltonian (i.e. $g^{\mu\nu}p_{\mu}p_{\nu}+m^2c^2$ becomes $g_{\mu\nu}\dot{q}^{\mu}\dot{q}^{\nu}+\hbar^2/m^2c^2$). If we perform first the integral over the world line metric, the mass shell constraint is implemented (see~\cite{CcapaTtira:2009ku} for a version implementing Dirichlet boundary conditions in field theory)
\vspace{0.5cm}
\begin{equation}
\label{eq:eintegral}
\fl \Delta(x,x)= \int^{\infty}_{0} d\tau \oint_{q[0]=q[1]=x} [dq]\sqrt{\det(g)}\int [dp/\hbar] \exp \left(-i/\hbar \int^{\tau}_{0}dt \dot{q}p\right)\mathbf{ \delta}(p^2+m^2c^2)
\end{equation}
\vspace{0.5cm}
(i.e the vanishing of the Hamiltonian on physical states) leaving the form of the reduced action operator encountered previously in Equation~(\ref{eq:Loop3}). Integrating over both the momenta and the world line metric would give us the nonlinear form only depending on the position
\begin{equation}
\label{eq:epintegral}
\Delta(x,x)= \int^{\infty}_{0} d\tau \oint_{q[0]=q[1]=x} [dq]\sqrt{\det(g)} \exp \left(-imc/\hbar \int^{\tau}_{0}dt \sqrt{g_{ij}\dot{q}^i\dot{q}^j}\right).
\end{equation}
The significance here is that the exponential contains both the path length of the loop and the perturbing potentials via the metric. Expanding the metric about the given background case, the perturbing potentials act as the scattering elements. The final schematic form then we take for the $\mathcal{Z}$-loop integral is
\vspace{0.5cm}
\begin{eqnarray}
\ \mathcal{Z}[\Omega,x] =&\langle\mathbf{F}_x | \int^{\infty}_{0} d\tau \oint_{q[0]=q[1]=x} [dq]\sqrt{\det(g)}\int [d\hat{D}] [de] \nonumber \\
&\cdot \exp \left(- \int^{\tau}_{0}dt[ \dot{q}\hat{D}+e(g^{ij}\hat{D}_i\hat{D}_j+1)]\right) |\mathbf{I}_x\rangle, 
\end{eqnarray}
where Equation~(\ref{eq:Loop3}) can then be considered a shorthand notation for this. The classical equations of motion then lead to closed geodesics defining the loops. In addition the integration over the frequency can provide the sum over all paths (i.e. a sum over all loop sizes where there is one geodesic for one particular frequency), whilst the tau integration once compactified gives the windings of the loops. For the scalar version given by Equation~(\ref{eq:twopointpi}) this is
\begin{eqnarray}
\ \mathcal{Z}[\Omega,x] =&\sum^{\infty}_{n=1} \langle\mathbf{F}_x | \int^{1}_{0} d\tau \oint_{q[0]=q[1]=x} [dq]\sqrt{\det [\Omega^2/c^2(\epsilon \cdot \mu)]}\int [d\hat{\nabla}] [de] \nonumber \\
&\cdot \exp \left(-n \int^{\tau}_{0}dt[ \dot{q} \cdot \hat{\nabla}+e(\hat{\nabla}_{\epsilon}\cdot \hat{\nabla}_{\mu}+1)]\right) |\mathbf{I}_x\rangle,
\end{eqnarray}
and
\begin{equation}
\int^{\infty}_{0}d\Omega \mathcal{Z}[\Omega,x] \sim \sum_{loops} \mathcal{Z}[\Omega,x].
\end{equation}
\section{Some Examples}
\label{sec:Examples}
\subsection{Electromagnetic fields in media}
In the case of electromagnetic fields propagating in a general background media, the loop integral is given by
\begin{equation}
\label{eq:LOOPEM}
\mathcal{Z}^1[\Omega,x]= \langle\mathbf{F}_x| \mathcal{P} \exp \left[-\oint_{\mathcal{C}_x}dq^i \hat{D}_j \right] |\mathbf{I}_x\rangle.
\end{equation}
To recover the previous results for the $N$-sphere scattering in Section~\ref{sec:Loop} , one can pass to the nonlinear form similar to Equation~(\ref{eq:epintegral}) and perform a perturbative analysis. The first step would be to expand the material properties around a  background value to linear order in the fluctuations
\begin{eqnarray}
\epsilon(q,\Omega) & = & \langle \epsilon_B(q,\Omega) \rangle +\delta \epsilon (q,\Omega),\\
\mu(q,\Omega) & = & \langle \mu_B(q,\Omega) \rangle+\delta \mu(q,\Omega). 
\end{eqnarray}
The fluctuations then describe the presence of each of the spheres, and the nonlinear form can be expanded (the square root of the corresponding line element). Following this it is necessary to introduce complete sets of vector eigenfunctions (spherical harmonics $|L \rangle$ and Bessel functions $|z_L \rangle$ that solve the Helmholtz equation for the constant background) centred on each of the spheres (coordinate system $\Sigma_i$) 
\begin{eqnarray}
\mathcal{H}&=&\{|TE_i \rangle =\mathbf{L}|L,z_L,\Sigma_i\rangle, |TM_i \rangle=\mathbf{\nabla}\wedge \mathbf{L}|L,z_L,\Sigma_i\rangle , \nonumber \\
& &  | LO_i \rangle =\mathbf{\nabla}|L,z_L,\Sigma_i\rangle \}.
\end{eqnarray}
Both the $\mathbf{E}$ and $\mathbf{B}$ are expanded in the complete set of states $\mathcal{H}$ with no longitudinal component to satisfy the divergence free equations of motion. The loop integral can be cut up into essentially all the path ways connecting the sphere centres. By inserting the complete sets of states we extract both the translation matrices already given in Equation~(\ref{eq:Arep}) and also the scattering coefficients (where $k=\sqrt{\epsilon_B\mu_B}\Omega /c$) given by
\begin{eqnarray}
\alpha^i_L & = & \frac{k\Omega^2}{2\pi c^2}\int_{B^2} \langle TM_i| \delta \epsilon  |TM_i \rangle, \\
\beta^i_L & = & \frac{k\Omega^2}{2\pi c^2}\int_{B^2} \langle TM_i | \delta \mu  |TM_i \rangle .
\end{eqnarray}
The canonical momentum density (that is consistent with the Lorentz force law)  is purely in the longitudinal direction $\mathbf{\nabla}$, which consists of the scattered modes with respect to sphere-$i$
\begin{eqnarray}
\mathbf{P}& = & \mathbf{E}^{\dagger}\wedge \mathbf{B}, \nonumber \\
& = &   \frac{i}{\omega}\mathbf{E}^{\dagger}\wedge \nabla \wedge \mathbf{E}, \nonumber \\
& = &   \frac{i}{2\omega}\mathbf{\nabla}(\mathbf{E}^{\dagger}\cdot  \mathbf{E}).
\end{eqnarray}
Since it is only the derivative operator entering here giving the direction of the momentum, this appears at least to be consistent with the simple derivative operator evaluated on the loop. The choice of physical momentum is a subtle point~\cite{PhysRevE.55.4207, Brevik1979133} (see also~\cite{tiggelen:130402,kawka-2009}) which we do not address in detail here.

\subsection{A scalar field in a potential}
As a toy model consider a pseudo-Hydrogen atom in two dimensions, with a scalar field  as the force mediator. That is we have two polarisable particles, one of which is big (the nucleus) compared to the other (the electron) that are chargeless and interacting through the dispersion force carrier (the scalar field). It should be stressed that whilst this is not a physical example, it can give some insight into handling singular potentials. The $\mathcal{Z}$-loop integral can be written as
\begin{equation}
\label{eq:LOOPSCALAR}
\mathcal{Z}^0[\Omega,x]= \sum_{\mathcal{C}_x}\langle\mathbf{F}_x| \mathcal{P} \exp \left[-\oint_{\mathcal{C}_x}dq^i \hat{D}_j \right] |\mathbf{I}_x\rangle.
\end{equation}
The system has a $U(1)$ symmetric potential $k^2(r,\Omega)$ with a length scale $R$ given by
\begin{equation}
\label{eq:potential}
k^2(r,\Omega)=\frac{\Omega^2 R}{c^2r}=\frac{\Omega^2 }{c^2}V(r),
\end{equation}
and is compatible with some two dimensional deformed metric
\begin{equation}
\label{eq:metric}
ds^2=V(r)(dr^2+r^2d\theta^2).
\end{equation}
The geodesic equations can be simplified using the constants of motion. Due to the $U(1)$ symmetry and the invariance of the action in the label $\tau$ we have
\begin{eqnarray}
\label{eq:amconstant}
\frac{d}{d\tau}(V r^2 \dot{\theta})=0, \\
\frac{d}{d\tau}(V [\dot{r}^2 +r^2\dot{\theta}^2])=0.
\end{eqnarray}
These can be rearranged such that they take the form one usually encounters in classical orbits,
\begin{equation}
\label{eq:ellipse}
\left[\frac{d^2u}{d\theta^2}+u\right]=\frac{E}{2}\frac{d V}{d u},
\end{equation}
where $u=1/r$ and $E$ is the integration constant associated with the constant energy orbits. Given the long range potential in Equation~(\ref{eq:potential}), it follows the loops are ellipses where the eccentricity $\mathcal{E}$ is determined by the frequency $\Omega$, the length scale $R$ and the integration constant $E$ (which can be chosen to be unity). The final sum then needs to be over all ellipses with the eccentricity being summed/integrated over, together with the winding number $n$ for how many times we wind around the ellipse
\begin{eqnarray}
\label{eq:LOOPSCALAR2}
\mathcal{Z}^0[\Omega,x]&=&\sum^{\infty}_{n=1} \sum_{\mathcal{E}}  \exp \left[-n\cdot l(x,\mathcal{E}) \right] \nonumber \\
 &=& \sum_{\mathcal{E}} \frac{ \exp[- l(x,\mathcal{E})]}{1- \exp[- l(x,\mathcal{E})]}.
\end{eqnarray}
Here $l$ is the length of the elliptical orbit that depends on the eccentricity and the point $x$ where the loop integral is being evaluated at (which determines the integration constants in the solution of  Equation~(\ref{eq:ellipse})).

\section{Conclusions}

The principal result of this paper is that we can represent Casimir interactions (which are given by bilinear correlation functions of the physical fields) in terms of an equivalent particle type path integral with an underlying loop structure . In passing from a field description to that of a particle description, we also find that the material properties (for example the permittivity and permeability) in which the fields propagate are turned into an appropriate geometric quantity (a connection or metric) in which the particle like quantity moves. By introducing the use of a connection to define parallel transport, it is possible to calculate the space of loops, which can be found from a simple geodesic equation. One can pass between different representations of the path integral by integrating over the worldline metrics and/or the conjugate translation operator (which is locally defined). From here we see how the loops come out from the second order form, how the nonlinear form can be used to evaluate expressions perturbatively, and how the first order Hamiltonian form has the translation operator contained within it. The final integral over frequency produces a sum over all the loops and in this way one finds a sum over all paths. It is expected to be principally useful when applied to Casimir physics, for those situations where the materials between bodies have real length scales associated with them; at certain frequencies one can imagine being able to resolve these length scales and the orbit sees an effective background curvature in which it is moving. To go beyond a straight line approximation in this situation is necessary and we expect the loop integrals presented here to be applicable.

It would be interesting in the future to investigate further the role of loop variables as the real dynamical variables from a loop quantum electrodynamics perspective~\cite{diBartolo:1983pt}. If we write the dynamical variables  explicitly in terms of the loops, we may well expect all the physical observables to be finite and not require the type of regularisation usually associated with quantum field theories. The inter body force is one such observable, but possibly one could apply it to constructions involving friction~\cite{scheel-2009} where one would have to use a theory of dynamic loops. In addition it would be interesting to see how the functional methods developed in~\cite{rahi-2009} could be related to the loop integral picture presented here. This will be left for future work.

\label{sec:Conclusions}

\ack
I wish to acknowledge Stefan Buhmann and John Gracey for useful discussions and critical remarks. Additionally I would like to thank Stefan Scheel and Alex Crosse for background comments. This work was supported by the SCALA programme of the European commission.

\section*{References}


\end{document}